\documentclass[a4paper,11pt]{article}
\pdfoutput=1 % if your are submitting a pdflatex (i.e. if you have
\usepackage{jheppub} % for details on the use of the package, please
\usepackage{slashed}

\usepackage[T1]{fontenc} % if needed

\title{\boldmath Nonrelativistic expansion of M2 branes and M theory backgrounds}

\author[a]{Dibakar Roychowdhury}
\affiliation[a]{Department of Physics, Indian Institute of Technology Roorkee,\\Roorkee 247667, Uttarakhand, India}
\emailAdd{dibakar.roychowdhury@ph.iitr.ac.in}
\abstract{We initiate a systematic analysis of the nonrelativistic membrane solutions of M theory using the notion of 11d membrane Newton-Cartan (MNC) geometry as well as considering a $ 1/c^2 $ expansion for the embedding fields of the M2 brane world-volume theory. We discuss the associated boost and dilatation symmetries of the nonrelativistic world-volume theory at leading order in the $ 1/c $ expansion. We show that, in the static gauge, when the world-volume directions of the nonrelativistic M2 brane are stretched along the longitudinal axes of the target space geometry, the leading order action in the $ 1/c $ expansion becomes trivial. In other words, the nontrivial dynamics appears to be only at NLO and beyond. In our analysis, we focus on such embeddings only and obtain the corresponding dispersion relation associated with the nonrelativistic world-volume theory. }
\begin{document} 
\maketitle
\flushbottom
\newpage
%%%%%%%%%%%%%%%%%%%%%%%%%%%%%%%%%%%%%%
\section{Introduction and summary}
Nonrelativistic (NR) string theory, that emerged due to the seminal work by Gomis and Ooguri \cite{Gomis:2000bd}-\cite{Danielsson:2000gi} has already started playing a pivotal role in our present day understanding of the theory of quantum gravity those are defined over Non-Lorentzian backgrounds. Starting from strings \cite{Gomis:2005pg}-\cite{Yan:2021lbe}, this formalism has now been developed for various extended objects \cite{Roychowdhury:2019qmp}-\cite{Bergshoeff:2022pzk} those play central role in a theory of NR quantum gravity.

The NR limit, that we consider in this paper, is achieved in two steps. The first step is to consider a $ p $ brane Newton-Cartan limit of the relativistic target space manifold \cite{Ebert:2021mfu} 
\begin{eqnarray}
g_{\mu \nu}&=&c^2 \tau_{\mu \nu}(x)+c^{1-p}H_{\mu \nu}(x)+ \cdots ,\\
C^{(p+1)}_{\mu \nu \cdots \lambda}&=&-c^{p+1}\epsilon_{AB \cdots C}\tau_{\mu}~^A (x) \tau_{\nu}~^B (x)\cdots \tau_{\lambda}~^C(x) +\hat{C}_{\mu \nu \cdots \lambda}(x)+\cdots ,
\end{eqnarray}
where $ \tau_{\mu \nu} $ is identified as the longitudinal metric associated with the Non-Lorentzian background. On the other hand, $ H_{\mu \nu} $ corresponds to the metric of the transverse space.

Clearly, the special case with $ p=1 $ corresponds to a String Newton-Cartan (SNC) limit of fundamental strings. However, for the purpose of this paper, we would focus on the M2 brane Newton-Cartan (MNC) limit of M theory which corresponds to setting $ p=2 $.

The second crucial ingredient of our NR construction is based upon a large $ c $ expansion of the world-volume fields for M2 branes those are defined over a $ p=2 $ MNC background. This formalism has recently been developed for closed strings propagating over SNC backgrounds \cite{Hartong:2021ekg} and is yet to be explored for extended objects like M2 branes\footnote{This is quite similar in spirit as that of a $ 1/c $ expansion of general relativity \cite{Hansen:2019pkl}.} \cite{Sezgin:2002rt}-\cite{Bozhilov:2005ew}.

When applied to closed strings, it turns out that the NR (with respect to the target space geometry) sigma model yields a consistent $ 1/c^2 $ expansion for the world-sheet fields in which case the divergence coming from the metric is precisely cancelled that due to the Kalb-Ramond fields. To summarise, one ends up in a large $ c $ expansion of the sigma model where the dynamics can be studied at each order in the $ 1/c^2 $ expansion. It therefore raises a natural question, whether such an expansion can also make sense for extended objects like M2 branes in M theory.

We build up the NR theory of M2 branes using the notion of the 11d membrane Newton-Cartan (MNC) geometry \cite{Blair:2021waq} together with an expansion of the embedding scalars (of the M2 brane world-volume theory) in a systematic $ 1/c^2 $ expansion \cite{Hartong:2021ekg}. When put together, these two expansions result in a nice $ 1/c $ expansion for the M2 brane world-volume theory where the leading order divergences are precisely cancelled due to the contributions coming from the background three form in the infinite speed limit of light.  

This new NR M2 brane world-volume action offers a natural playground for further investigations. In the present paper, we explore some of these aspects and leave the rest of the questions as a part of future investigations.

Our analysis reveals that the leading order (LO) Lagrangian density vanishes in the \emph{static} gauge when the world-volume directions are stretched along the longitudinal axes of the MNC manifold. We verify this claim both for flat as well as curved MNC backgrounds. In other words, the dynamics in such cases commences only at NLO and beyond.  

The organisation for the rest of the paper is as follows. In Section \ref{sec2}, we elaborate on the $ 1/c^2 $ formalism \cite{Hartong:2021ekg} for the M2 branes those are defined over MNC backgrounds. We show that the MNC expansion(s) together with a $ 1/c^2 $ expansion for the world-volume degrees of freedom leads to a consistent NR expansion for the M2 brane world-volume theory. We further explore the symmetries of the this NR theory at LO in the $ 1/c $ expansion. 

In Section \ref{sec3}, we explore the NR world-volume theory for flat MNC backgrounds. It turns out that at NLO, the world-volume theory is described by the transverse fluctuations ($ x^I $) of the brane those are at leading order in the $ 1/c^2 $ expansion of  (\ref{2.10}). At NNLO, the world-volume theory is solely determined by the longitudinal fluctuations ($ y^{\hat{A}} $) those are at next to leading order in the $ 1/c^2 $ expansion (\ref{2.10}).

In Section \ref{sec4}, the analysis is further generalised for M2 branes propagating over $ AdS_4 \times S^7 $. To simplify our analysis, we confine the M2 brane within $ AdS_4 $ in which case the longitudinal axes of the MNC target space are extended along the world-volume directions of the brane and all the leading order transverse fluctuations are switched off.

We obtain the corresponding MNC data and show that the nontrivial dynamics commences only at NNLO in the $ 1/c $ expansion. We obtain the dynamics associated to longitudinal fluctuations (associated to the radial direction of the $ AdS_4 $) at next to leading order in the expansion (\ref{2.10}) and find a consistent solution for the \emph{folded} M2 brane configurations.

In Section \ref{sec5}, we generalize our results by incorporating spin ($ S $) for NR folded M2 brane configurations. The spin corresponds to a frequency of rotation along one of the compact isometry direction of the MNC target space. Our analysis reveals a dispersion relation (between the energy ($ E_{NR} $) and the spin ($ S $) of the NR membrane) of the form 
\begin{eqnarray}
E_{NR}\sim S^2.
\end{eqnarray}

Finally, we draw our conclusion in Section \ref{sec6}, where we list a set of problems those might be pursued in the near future and outline some basic steps related to that.
%%%%%%%%%%%%%%%%%%%%%%%%%%%%%%%%%%%%%%
\section{Membrane Newton-Cartan geometry and M2 branes}
\label{sec2}
Relativistic M2 branes are described by an world-volume action of the following form \cite{Alishahiha:2002sy}
\begin{eqnarray}
\label{2.1}
S_{M2}&=&T_2 \int d^3 \sigma \sqrt{- \det g_{mn}}+T_2 \int C^{(3)},\nonumber\\
&=& S_{NG}+S_{WZ}.
\end{eqnarray}

Here, $ X^{\mu} ~(\mu=0, \cdots ,10) $ are the target space directions and $ m,n =0,1,2 $ are the world-volume directions. On the other hand, $ C^{(3)} $ is the background three form flux (to which the M2 brane is coupled to) together with
\begin{eqnarray}
g_{mn}=g_{\mu \nu}(X)\partial_m X^{\mu} \partial_n X^{\nu},
\end{eqnarray}
as the induced metric on the world-volume of the M2 brane.
%%%%%%%%%%%%%%%%%%%%%%%%%%%%%%%%%%%%%%
\subsection{Membrane Newton-Cartan data}
Our analysis closely follows the algorithm developed in \cite{Blair:2021waq}. Like in a standard NR approach, we split the geometry into three longitudinal ($ A=0,1,2 $) and eight transverse ($ a=3, \cdots , 10 $) directions. In what follows, we consider the M2 brane world-volume directions are stretched along the longitudinal axes while the remaining eight directions correspond to the transverse fluctuations of the brane.

The eleven dimensional metric and its inverse is expressed as\footnote{Here, the metric components $ \tau_{\mu \nu} $ and $ H_{\mu \nu} $ can be further Taylor expanded considering the expansion (\ref{2.10}). These are the expansions of the form $ \tau_{\mu \nu}=\tau_{\mu \nu}(x)+c^{-2}y^\lambda \partial_\lambda \tau_{\mu \nu}+\cdots $ and $ H_{\mu \nu}= H_{\mu \nu}(x)+c^{-2}y^\lambda \partial_\lambda H_{\mu \nu}+\cdots$. This gives rise to a zeroth order term which is disregarded considering the fact that the longitudinal vielbein is a \emph{slowly} varying function of the target space cordinates. On the other hand, the expansion of $ H_{\mu \nu} $ gives rise to a $ \mathcal{O}(c^{-3}) $ term which can be ignored for the purpose of the present analysis. Similar remarks also hold for each term in the expansion (\ref{2.8}) where for the three form flux field $ \hat{C}_{\mu \nu \lambda} $ we ignore its spacetime variations considering it to be a slowly varying function of the target space coordinates.} \cite{Ebert:2021mfu}-\cite{Blair:2021waq}
\begin{eqnarray}
g_{\mu \nu} &=& c^2 \tau_{\mu \nu}(x)+c^{-1}H_{\mu \nu}(x)+\cdots ,\\
g^{\mu \nu} &=& c H^{\mu \nu}(x)+c^{-2} \tau^{\mu \nu}(x)+\cdots ,
\end{eqnarray}
where we restrict ourselves upto $ \mathcal{O}(c^{-2}) $ and denote 
\begin{eqnarray}
 \tau_{\mu \nu} =\tau_{\mu}~^A \tau_{\nu}~^B  \eta_{A B}~;~H_{\mu \nu}=h_{\mu}~^a h_{\nu}~^b \delta_{ab}.
\end{eqnarray}

Here $ \tau_{\mu}^A $ are the clock one forms and $ H_{\mu \nu} $ is the metric of the transverse manifold together with their inverses defined as
\begin{eqnarray}
\tau^{\mu \nu}&=&\tau^{\mu}~_A \tau^{\nu}~_B \eta^{AB}~;~H^{\mu \nu}=h^{\mu}~_a h^{\nu}~_b \delta^{ab},\\
\tau^{\mu}~_A \tau_{\mu}~^B &=&\delta^B_A ~;~h^{\mu}~_a h_{\mu}~^b=\delta^b_a .
\end{eqnarray}

The three form flux, on the other hand, is expanded as \cite{Blair:2021waq}
\begin{eqnarray}
\label{2.8}
C_{\mu \nu \lambda}=-c^3 \epsilon_{ABC}\tau_{\mu}~^A \tau_{\nu}~^B \tau_{\lambda}~^C +\hat{C}_{\mu \nu \lambda}(x)+\cdots .
\end{eqnarray}

Here, $ c $ is the speed of light and the NR limit corresponds to setting, $ c \rightarrow \infty $. The collective background fields $ \lbrace \tau_{\mu}~^A, H_{\mu \nu}, \hat{C}_{\mu \nu \lambda}, \tilde{C}_{\mu \nu \lambda} \rbrace $ is what we define as the M2 brane Newton-Cartan (MNC) data in this paper.

Following the normalisation condition for the metric namely, $ g_{\mu \rho}g^{\rho \nu}=\delta^{\nu}_{\mu} $ one finds
\begin{eqnarray}
\label{2.9}
\tau_{\mu \rho}H^{\rho \nu}=0~;~\tau_{\mu \rho}\tau^{\rho \nu}+H_{\mu \rho}H^{\rho \nu}=\delta^{\nu}_{\mu}  ~;~H_{\mu \rho}\tau^{\rho \nu}=0.
\end{eqnarray}
%%%%%%%%%%%%%%%%%%%%%%%%%%%%%%%%%%%%%%
\subsection{Expanding the world-volume action}
In order to obtain the NR world-volume theory for M2 branes, in addition to the above MNC data, one has to consider a large $ c $ expansion of the embedding fields \cite{Hartong:2021ekg}
\begin{eqnarray}
\label{2.10}
X^{\mu}=x^{\mu}+c^{-2}y^{\mu}+\mathcal{O}(c^{-4}),
\end{eqnarray}
where $ x^{\mu} $ stands for the leading order (LO) fluctuations and $ y^{\mu} $ as NLO fluctuations. 

Using the MNC data and the expansion (\ref{2.10}), we compute the following
\begin{eqnarray}
\label{2.11}
g_{mn}=c^2 \tau^{(2)}_{mn}+\tau^{(0)}_{mn}+c^{-1}H^{(-1)}_{mn}+c^{-2}\tau^{(-2)}_{mn}+\mathcal{O}(c^{-3}),
\end{eqnarray}
where we define the above entities as
\begin{eqnarray}
\tau^{(2)}_{mn}&=&\tau_{\mu \nu}\partial_{m}x^{\mu}\partial_n x^{\nu},\\
\tau^{(0)}_{mn}&=&\tau_{\mu \nu}(\partial_{m}x^{\mu}\partial_n y^{\nu}+\partial_m y^\mu \partial_n x^\nu),\\
\tau^{(-2)}_{mn}&=&\tau_{\mu \nu}\partial_{m}y^{\mu}\partial_n y^{\nu},\\
H^{(-1)}_{mn}&=& H_{\mu \nu}\partial_{m}x^{\mu}\partial_{n}x^{\nu}.
\end{eqnarray}

Using (\ref{2.11}) and introducing the inverse $ 3 \times 3 $ matrices $ \tilde{\tau}^{mn(2)} $ such that $ \tau_{mn}^{(2)}\tilde{\tau}^{nk(2)}=\delta^k_m $ one can expand the determinant in the large $ c $ limit as
\begin{eqnarray}
\label{2.16}
\sqrt{- \det g_{mn}}=c^3 \sqrt{- \det \tau^{(2)}_{mn}}~\mathcal{F}(\tau_{mn}, H_{mn}),
\end{eqnarray}
where we define the function in the large $ c $ limit as
\begin{align}
\mathcal{F}(\tau_{mn}, H_{mn}) &= 1+\frac{c^{-2}}{2}\tilde{\tau}^{mn(2)} \tau_{mn}^{(0)}+\frac{c^{-3}}{2}\tilde{\tau}^{mn(2)} H_{mn}^{(-1)}\nonumber\\
&+\frac{c^{-4}}{2}\tilde{\tau}^{mn(2)} \tau_{mn}^{(-2)} + \mathcal{O}(c^{-5}) .
\end{align}

Let us now focus on the WZ term in (\ref{2.1}). As a first step, we find
\begin{eqnarray}
\label{2.18}
\frac{1}{6}\epsilon^{mnp}\partial_m X^{\mu} \partial_n X^{\nu} \partial_p X^{\lambda} C_{\mu \nu \lambda}=-c^3 \det (\tau^{(2)})_m^A +c h^{(1)} +h^{(0)}\nonumber\\
+c^{-1}h^{(-1)}+c^{-2}h^{(-2)}+\mathcal{O}(c^{-3}),
\end{eqnarray}
where we identify the above functions as
\begin{eqnarray}
h^{(0)} &=& \frac{\epsilon^{mnp}}{6}\partial_m x^{\mu} \partial_n x^{\nu} \partial_p x^{\lambda} \hat{C}_{\mu \nu \lambda},\\
h^{(1)} &=&- \frac{\epsilon^{mnp}}{2}\epsilon_{ABC}\tau_{\mu}~^A \tau_{\nu}~^B \tau_{\lambda}~^C \partial_p y^{\lambda}\partial_m x^{\mu}\partial_n x^{\nu} ,\\
h^{(-1)} &=&- \frac{\epsilon^{mnp}}{2}\epsilon_{ABC}\tau_{\mu}~^A \tau_{\nu}~^B \tau_{\lambda}~^C \partial_p x^{\lambda}\partial_m y^{\mu}\partial_n y^{\nu} , \\
h^{(-2)} &=& \frac{\epsilon^{mnp}}{2}\hat{C}_{\mu \nu \lambda}\partial_p y^{\lambda}\partial_m x^{\mu}\partial_n x^{\nu}.
\label{2.24}
\end{eqnarray}

Adding (\ref{2.16}) and (\ref{2.18}) together and considering the fact 
\begin{eqnarray}
\sqrt{- \det \tau^{(2)}_{mn}} = \det (\tau^{(2)})_m^A \equiv \frac{\epsilon^{mnl}}{3!}\epsilon_{ABC}\tau_m ~^A \tau_n ~^B \tau_l ~^C,
\end{eqnarray}
we find the corresponding Lagrangian density in the large $ c $ limit as
\begin{align}
\label{2.25}
c^{-1}\mathcal{L}_{M2}&= \left(\frac{\sqrt{- \det \tau^{(2)}_{mn}}}{2}\tilde{\tau}^{mn(2)} \tau_{mn}^{(0)} +h^{(1)} \right) +c^{-1}\left(\frac{\sqrt{- \det \tau^{(2)}_{mn}}}{2}\tilde{\tau}^{mn(2)} H_{mn}^{(-1)} +h^{(0)}\right) \nonumber\\
&+c^{-2}\left( \frac{\sqrt{- \det \tau^{(2)}_{mn}}}{2}\tilde{\tau}^{mn(2)} \tau_{mn}^{(-2)} +h^{(-1)}\right).
\end{align}

The above expression (\ref{2.25}) could be schematically expressed as
\begin{eqnarray}
\label{2.26}
c^{-1}\mathcal{L}_{M2}= \tilde{\mathcal{L}}^{(0)}+c^{-1}\tilde{\mathcal{L}}^{(-1)}+c^{-2}\tilde{\mathcal{L}}^{(-2)}+ \cdots .
\end{eqnarray}

Introducing the M2 brane tension in the NR limit as
\begin{eqnarray}
\tilde{T}_2 = c T_2,
\end{eqnarray}
one obtains a NR expansion of the M2 brane world-volume action of the form
\begin{eqnarray}
\label{2.28}
S_{M2}&=&T_{2}\int \mathcal{L}_{M2}=\tilde{T}_{2}\int (\tilde{\mathcal{L}}^{(0)}+c^{-1}\tilde{\mathcal{L}}^{(-1)}+c^{-2}\tilde{\mathcal{L}}^{(-2)}),\nonumber\\
&=& \tilde{\mathcal{S}}^{(0)}+c^{-1}\tilde{\mathcal{S}}^{(-1)}+c^{-2}\tilde{\mathcal{S}}^{(-2)}+\cdots .
\end{eqnarray}
%%%%%%%%%%%%%%%%%%%%%%%%%%%%%%%%%%%%%%
\subsection{Symmetries at LO}
%%%%%%%%%%%%%%%%%%%%%%%%%
\subsubsection{Rotation}
Let us discuss the symmetries of the NR M2 brane action (\ref{2.28}). Clearly, (\ref{2.28}) is manifestly invariant under the world-volume diffeomorphisms. On top of this, it is also invariant under a global $ SO(8) $ rotation that rotates traverse ($ a =3, \cdots , 10$) indices 
\begin{eqnarray}
h_\mu ~^a =\Lambda_b~^a h_{\mu}~^b,
\end{eqnarray}
and thereby leaving $ H_{\mu \nu} $ invariant.

On a similar note, longitudinal ($ A=0,1,2 $) indices are rotated under $ SO(1,2) $ rotation
\begin{eqnarray}
\tau_\mu ~^A = \Lambda _B~^A \tau_\mu~ ^B,
\end{eqnarray}
which also leaves the NR action (\ref{2.28}) invariant.
%%%%%%%%%%%%%%%%%%%%%%%%%
\subsubsection{Galilean boost}
We now focus on Galilean boosts those mix longitudinal and transverse indices. Boost matrices are denoted as $ \Lambda_a~^A $ which result in an infinitesimal transformation of the form 
\begin{eqnarray}
\label{2.32}
\delta_\Lambda H_{\mu \nu}&=&2\Lambda_{(\mu}~^A \tau_{\nu) A} ,\\
\delta_\Lambda \tau^\mu~_A &=& -H^{\mu \nu}\Lambda_{\nu A},
\label{2.34}
\end{eqnarray}
where the boost matrix $ \Lambda_a~^A $ is related to $ \Lambda_\mu~^A $ through the transverse vielbein as \cite{Blair:2021waq}
\begin{eqnarray}
 \Lambda_\mu~^A = h^a~_\mu \Lambda_a~^A ,
\end{eqnarray}
subjected to the fact that, $ \tau^{\mu}~_A \Lambda_{\mu}~^B =0 $.

Using the above set of transformation rules (\ref{2.32})-(\ref{2.34}), one finds different expressions at different orders in the large $ c $ expansion. Let us explore the $ \mathcal{O}(c^0) $ term in the expansion
\begin{eqnarray}
\label{2.36}
\delta_{\Lambda}\tilde{\mathcal{L}}^{(0)} =\frac{1}{2} (\det (\tau^{(2)})_m^A)(\delta_{\Lambda}\tilde{\tau}^{mn (2)}\tau_{mn}^{(0)}+\tilde{\tau}^{mn (2)}\delta_{\Lambda}\tau_{mn}^{(0)})\nonumber\\+\frac{1}{2}\delta_{\Lambda} (\det (\tau^{(2)})_m^A)\tilde{\tau}^{mn (2)}\tau_{mn}^{(0)} +\delta_{\Lambda}h^{(1)}.
\end{eqnarray}

Let us compute each term in (\ref{2.36}) separately. A straightforward computation reveals
\begin{eqnarray}
\label{2.37}
\delta_{\Lambda} (\det (\tau^{(2)})_m^A)=\frac{\epsilon^{mnl}}{2}\epsilon_{ABC}  \tau_n ~^{(2)B} \tau_l ~^{(2)C} (\delta_{\Lambda} \tau_\mu ~^A \partial_m x^\mu + \tau_\mu ~^A \delta_{\Lambda} (\partial_m x^\mu)),
\end{eqnarray}
where we denote $ \tau_m ~^{(2)A}=\tau_\mu ~^A \partial_m x^\mu $.

Below, we further simplify (\ref{2.37}) using the following identity
\begin{eqnarray}
\delta_{\Lambda}\tau^\mu ~_A \tau_\mu ~^A+\tau^\mu ~_A \delta_{\Lambda} \tau_\mu ~^A =0.
\end{eqnarray}

After further simplification, one finds
\begin{eqnarray}
\tau^\mu ~_A \delta_{\Lambda} \tau_\mu ~^A =H^{\mu \nu}\Lambda_{\nu A}\tau_\mu ~^A,
\end{eqnarray}
which clearly reveals a solution of the form
\begin{eqnarray}
\delta_{\Lambda} \tau_\mu ~^A = \tau_\mu ~^B H^{\nu \rho}\Lambda_{\rho B}\tau_\nu ~^A .
\end{eqnarray}

On the other hand, after some algebra one finds the following variation
\begin{eqnarray}
\delta_{\Lambda}h^{(1)} = -\epsilon^{mnl}\epsilon_{ABC}\tau_n ~^{(2)B} \tau_l ~^{(-2)C}(\delta_{\Lambda} \tau_\mu ~^A \partial_m x^\mu + \tau_\mu ~^A \delta_{\Lambda} (\partial_m x^\mu))\nonumber\\
-\frac{\epsilon^{mnl}}{2}\epsilon_{ABC} \tau_n ~^{(2)B}\tau_l ~^{(2)C}(\delta_{\Lambda} \tau_\mu ~^A \partial_m y^\mu + \tau_\mu ~^A \delta_{\Lambda} (\partial_m y^\mu)),
\end{eqnarray}
where we define $ \tau_m ~^{(-2)A}=\tau_\mu ~^A \partial_m y^\mu $.

Finally, we note down the variation
\begin{eqnarray}
\frac{1}{2}\tilde{\tau}^{mn (2)}\delta_{\Lambda}\tau_{mn}^{(0)}=\tilde{\tau}^{mn (2)}\tau^{(-2)}_{nA}(\delta_{\Lambda} \tau_\mu ~^A \partial_m x^\mu + \tau_\mu ~^A \delta_{\Lambda} (\partial_m x^\mu))\nonumber\\
+\tilde{\tau}^{mn (2)}\tau^{(2)}_{nA}(\delta_{\Lambda} \tau_\mu ~^A \partial_m y^\mu + \tau_\mu ~^A \delta_{\Lambda} (\partial_m y^\mu)).
\end{eqnarray}

Clearly, the $ \mathcal{O}(c^0) $ world-volume theory is invariant under Galilean boost if we impose
\begin{eqnarray}
\label{2.43}
\delta_{\Lambda} \tau_\mu ~^A \partial_m x^\mu + \tau_\mu ~^A \delta_{\Lambda} (\partial_m x^\mu) &=&0,\\
\delta_{\Lambda} \tau_\mu ~^A \partial_m y^\mu + \tau_\mu ~^A \delta_{\Lambda} (\partial_m y^\mu)&=&0.
\label{2.44}
\end{eqnarray}

The above set of equations (\ref{2.43})-(\ref{2.44}) could be combined together to obtain
\begin{eqnarray}
\label{2.45}
\tau_\mu ~^A \delta_{\Lambda} (\partial_m X^\mu ) &=& -\tau_\mu ~^B H^{\nu \rho}\Lambda_{\rho B}\tau_\nu ~^A \partial_m X^\mu \nonumber\\
&=& -\tau_\nu ~^B H^{\mu \rho}\Lambda_{\rho B}\tau_\mu ~^A \partial_m X^\nu.
\end{eqnarray}

The above relation (\ref{2.45}) yields the following transformation under Galilean boost
\begin{eqnarray}
\label{2.46}
\delta_{\Lambda} (\partial_m X^\mu ) = -\tau_\nu ~^B H^{\mu \rho}\Lambda_{\rho B} \partial_m X^\nu \equiv -\tau_m ~^B H^{\mu \rho}\Lambda_{\rho B},
\end{eqnarray}
which we identify as the transformation of the gradient of a world-volume field under the Galilean boost such that the leading order action is boost invariant.

Assuming that the variation mutually commutes with the derivative operations, one could further integrate (\ref{2.46}) partially to obtain
\begin{eqnarray}
\delta_{\Lambda} X^{\mu}= -\tau_\nu ~^B H^{\mu \rho}\Lambda_{\rho B} X^\nu + \int \partial_m (\tau_\nu ~^B H^{\mu \rho}\Lambda_{\rho B})X^\nu d\sigma^m .
\end{eqnarray}

Using the above constraint (\ref{2.43}), it is also straightforward to show
\begin{eqnarray}
\label{2.47}
\tau_{mn}^{(2)} \delta_{\Lambda} \tilde{\tau}^{mn (2)}&=&-\tilde{\tau}^{mn (2)}\delta_{\Lambda}\tau_{mn}^{(2)}\nonumber\\
&=&-2\tilde{\tau}^{mn (2)}\tau^{(2)}_{nA}(\delta_{\Lambda} \tau_\mu ~^A \partial_m x^\mu + \tau_\mu ~^A \delta_{\Lambda} (\partial_m x^\mu))=0.
\end{eqnarray}

Since, $ \tau_{mn}^{(2)} \neq 0 $, therefore the only way to satisfy the above condition (\ref{2.47}) is to demand that $  \delta_{\Lambda} \tilde{\tau}^{mn (2)} =0 $ under the Galilean boost. Putting all these pieces together, one ensures the invariance of the zeroth order Lagrangian ($ \tilde{\mathcal{L}}^{(0)} $) under the Galilean boost.
%%%%%%%%%%%%%%%%%%%%%%%%%
\subsubsection{Dilatations}
Let us now explore the dilatation symmetries of the NR action (\ref{2.28}) at leading order in the $ 1/c $ expansion. The dilation transformations are introduced as 
\begin{eqnarray}
\label{2.48}
\delta_{\lambda}H^{\mu \nu}=-2\lambda H^{\mu \nu}~;~\delta_{\lambda}H_{\mu \nu}=2\lambda H_{\mu \nu}~;~\delta_{\lambda} \tau^{\mu}~_A = -\lambda \tau^{\mu}~_A ~;~\delta_{\lambda} \tau_{\mu}~^A = \lambda \tau_{\mu}~^A.
\end{eqnarray}

Under the action of (\ref{2.48}), the zeroth order Lagrangian transforms as
\begin{eqnarray}
\label{2.49}
\delta_{\lambda}\tilde{\mathcal{L}}^{(0)} =\frac{1}{2} (\det (\tau^{(2)})_m^A)(\delta_{\lambda}\tilde{\tau}^{mn (2)}\tau_{mn}^{(0)}+\tilde{\tau}^{mn (2)}\delta_{\lambda}\tau_{mn}^{(0)})\nonumber\\+\frac{1}{2}\delta_{\lambda} (\det (\tau^{(2)})_m^A)\tilde{\tau}^{mn (2)}\tau_{mn}^{(0)} +\delta_{\lambda}h^{(1)}.
\end{eqnarray}

Below, we enumerate each of the individual terms in (\ref{2.49}) separately. 
\begin{eqnarray}
\delta_{\lambda} (\det (\tau^{(2)})_m^A)=\frac{\epsilon^{mnl}}{2}\epsilon_{ABC}  \tau_n ~^{(2)B} \tau_l ~^{(2)C} (\lambda \tau_\mu ~^A \partial_m x^\mu + \tau_\mu ~^A \delta_{\lambda} (\partial_m x^\mu)),
\end{eqnarray}
\begin{eqnarray}
\delta_{\lambda}h^{(1)} = -\epsilon^{mnl}\epsilon_{ABC}\tau_n ~^{(2)B} \tau_l ~^{(-2)C}(\lambda \tau_\mu ~^A \partial_m x^\mu + \tau_\mu ~^A \delta_{\lambda} (\partial_m x^\mu))\nonumber\\
-\frac{\epsilon^{mnl}}{2}\epsilon_{ABC} \tau_n ~^{(2)B}\tau_l ~^{(2)C}(\lambda \tau_\mu ~^A \partial_m y^\mu + \tau_\mu ~^A \delta_{\lambda} (\partial_m y^\mu)),
\end{eqnarray}
\begin{eqnarray}
\tau_{mn}^{(2)} \delta_{\lambda} \tilde{\tau}^{mn (2)}&=&-\tilde{\tau}^{mn (2)}\delta_{\lambda}\tau_{mn}^{(2)}\nonumber\\
&=&-2\tilde{\tau}^{mn (2)}\tau^{(2)}_{nA}(\delta_{\lambda} \tau_\mu ~^A \partial_m x^\mu + \tau_\mu ~^A \delta_{\lambda} (\partial_m x^\mu)).
\end{eqnarray}

Therefore, the leading order world-volume theory is invariant under the action of dilatation (\ref{2.48}) provided the world-volume scalar transforms as
\begin{eqnarray}
\label{2.52}
\delta_{\lambda}(\partial_m x^{\mu})=-\lambda  \partial_m x^{\mu}.
\end{eqnarray} 
This further sets, $ \delta_{\lambda} \tilde{\tau}^{mn (2)}=0 $ and $ \delta_{\lambda} (\det (\tau^{(2)})_m^A)=0 $ under dilatations.

Assuming that the operations $\delta_{\lambda}  $ and $ \partial_m $ mutually commute, the above relation (\ref{2.52}) leads to a simple scaling relation for the world-volume fields under dilatation
\begin{eqnarray}
\delta_{\lambda} x^{\mu}=-\lambda x^{\mu}.
\end{eqnarray}
%%%%%%%%%%%%%%%%%%%%%%%%%%%%%%%%
\subsection{Symmetries at NLO}
We now explore the symmetries of the NR world-volume theory at NLO. In what follows, we use the results of the previous section to derive transformation properties of the world-volume fields those are pertinent to the NLO action.

A straightforward calculation shows that under the Galilean boost
\begin{align}
\label{ee2.56}
\delta_{\Lambda}\tilde{\mathcal{L}}^{(-1)} &= (\det (\tau^{(2)})_m^A)\tilde{\tau}^{mn(2)}\left(\Lambda_{(\mu}~^A \tau_{\nu)A} -H_{\lambda \nu}H^{\lambda \rho}\tau_\mu~^B\Lambda_{\rho B}\right) \partial_m x^{\mu} \partial_n x^{\nu}\nonumber\\
&+\frac{\epsilon^{mnp}}{6}\partial_m x^{\mu} \partial_n x^{\nu} \partial_p x^{\lambda} \Big(  \delta_{\Lambda}\hat{C}_{\mu \nu \lambda}-3\tau_\mu~^B H^{\beta \rho}\Lambda_{\rho B}\hat{C}_{\beta \nu \lambda}\Big).
\end{align}

Using the identity (\ref{2.9}) and considering the fact $ \tau^\rho~_C \Lambda_{\rho B}=0 $ \cite{Blair:2021waq}, it is quite straightforward to show the vanishing of the first term in (\ref{ee2.56})
\begin{align}
\tilde{\tau}^{mn(2)}\left(\Lambda_{(\mu}~^A \tau_{\nu)A} -H_{\lambda \nu}H^{\lambda \rho}\tau_\mu~^B\Lambda_{\rho B}\right) \partial_m x^{\mu} \partial_n x^{\nu}=0.
\end{align}

Therefore, one is left with the second term in (\ref{ee2.56}) vanishing of which imposes the following constraint on the world-volume fields under Galilean boost
\begin{align}
\delta_{\Lambda}\hat{C}_{\mu \nu \lambda}+3\tau_\mu~^A \hat{C}_{\rho \nu \lambda}\delta_{\Lambda}\tau^\rho~_A =0.
\end{align}

A straightforward calculation shows that under local scaling of fields
\begin{align}
\label{ee2.59}
\delta_{\lambda}\tilde{\mathcal{L}}^{(-1)} &= (\det (\tau^{(2)})_m^A)\tilde{\tau}^{mn(2)}\left( \delta_\lambda H_{\mu \nu}\partial_m x^\mu \partial_n x^\nu +2 H_{\mu \nu}\delta_\lambda (\partial_m x^\mu)\partial_n x^\nu\right) \nonumber\\
&+\frac{\epsilon^{mnp}}{6}\partial_m x^{\mu} \partial_n x^{\nu} \partial_p x^{\lambda} \left(\delta_\lambda \hat{C}_{\mu \nu \lambda}-3 \lambda \hat{C}_{\mu \nu \lambda} \right). 
\end{align}

The first term in (\ref{ee2.59}) vanishes identically by virtue of (\ref{2.48}) and (\ref{2.52}). On the other hand, the vanishing of the second term in (\ref{ee2.59}) fixes the transformation properties of the three form fluxes under dilatations as
\begin{align}
\delta_\lambda \hat{C}_{\mu \nu \lambda}=3 \lambda \hat{C}_{\mu \nu \lambda}.
\end{align}
%%%%%%%%%%%%%%%%%%%%%%%%%%%%%%%%%%%%%%
\section{Nonrelativistic M2 branes in flat MNC background}
\label{sec3}
%%%%%%%%%%%%%%%%%%%%%%%%%%%%%%%%%%
\subsection{MNC data}
As a special case, we consider M2 brane dynamics in \emph{flat} target space
\begin{eqnarray}
\label{3.1}
\tau_{\mu}~^0 = \delta_{\mu}^t~;~ \tau_{\mu}~^1 = \delta_{\mu}^u~;~\tau_{\mu}~^2 = \delta_{\mu}^v~;~H_{\mu \nu}=\delta_{\mu}^I \delta_{\nu}^I,
\end{eqnarray}
where we identify $ \lbrace t, u ,v\rbrace $ as longitudinal directions while $ \lbrace I, J\rbrace $ being the transverse indices associated with the flat MNC background.

Using (\ref{3.1}), one obtains the following data associated with the flat MNC target space
\begin{eqnarray}
\tau^{(2)}_{mn}&=&-\partial_{m}x^{t}\partial_n x^{t}+\partial_{m}x^{a}\partial_n x^{a}\\\tau^{(-2)}_{mn}&=&-\partial_{m}y^{t}\partial_n y^{t}+\partial_{m}y^{a}\partial_n y^{a}~;~
H^{(-1)}_{mn}= \partial_{m}x^{I}\partial_{n}x^{I},\\
\tau^{(0)}_{mn}&=&-(\partial_{m}x^{t}\partial_n y^{t}+\partial_m y^t \partial_n x^t)+\partial_{m}x^{a}\partial_n y^{a}+\partial_m y^a \partial_n x^a,
\end{eqnarray}
where $ a=u,v $ stands for the longitudinal axes.

In order to find the inverse $ \tilde{\tau}^{mn(2)} $ we choose to work with the following ansatz\footnote{The picture that we have in mind is that of a NR M2 brane extended along the longitudinal axes ($ t ,u ,v $) while $ I,J $ being the directions transverse to the brane.} for the \emph{longitudinal} world-volume fields namely
\begin{eqnarray}
\label{3.6}
x^t = x^t (\sigma^0)~;~x^u = x^u (\sigma^1)~;~x^v=x^v(\sigma^2),
\end{eqnarray}
where $ \lbrace \sigma^0 , \sigma^1 , \sigma^2 \rbrace = \lbrace \sigma^0 , \sigma^i \rbrace $ are the M2 brane world-volume directions.

This leads to the following matrix elements
\begin{eqnarray}
\tau^{(2)}_{00}=- (\dot{x}^t)^2~;~\tau^{(2)}_{0i}=\tau^{(2)}_{i0}=0~;~\tau^{(2)}_{ij}=(x'^a)^2,~i,j =1,2
\end{eqnarray}
where we define $ \dot{x}^t = \partial_0 x^t $ and $ x'^a = \partial_i x^a $.

Clearly, with the above choice (\ref{3.6}), the inverse matrix elements are quite straightforward to note down
\begin{eqnarray}
\tilde{\tau}^{mn(2)}= (\tau^{(2)}_{mn})^{-1}.
\end{eqnarray}
%%%%%%%%%%%%%%%%%%%%%%%%%%%%%%%%%%%%%%
\subsection{Dynamics at LO}
We begin by exploring the LO dynamics of NR M2 branes in an $ 1/c $ expansion. The leading order Lagrangian (density) is given by
\begin{eqnarray}
\tilde{\mathcal{L}}^{(0)}=\frac{1}{2}\sqrt{-\det \tau_{mn}^{(2)}}\tilde{\tau}^{mn(2)}\tau_{mn}^{(0)}+h^{(1)},
\end{eqnarray}
where a straightforward computation reveals
\begin{eqnarray}
\label{3.10}
\frac{1}{2}\sqrt{-\det \tau_{mn}^{(2)}}\tilde{\tau}^{mn(2)}\tau_{mn}^{(0)}=\dot{y}^t x'^u x'^v + \dot{x}^t (x'^v y'^u +x'^u y'^v),
\end{eqnarray}
together with
\begin{eqnarray}
\label{3.11}
h^{(1)} = -\dot{x}^t (x'^u y'^v +x'^v y'^u)-\dot{y}^t x'^v x'^u 
\end{eqnarray}
where $ y'^u = \partial_1 y^u $ and $ y'^v = \partial_2 y^v $.

Combining (\ref{3.10}) and (\ref{3.11}), the leading order Lagrangian vanishes identically
\begin{eqnarray}
\label{3.12}
\tilde{\mathcal{L}}^{(0)}=0.
\end{eqnarray}

Therefore, we conclude that nothing exactly happens at (static gauge) leading order in the $ 1/c $ expansion. The vanishing of the leading order theory (\ref{3.12}) is a consequence of the M2 brane embedding (\ref{3.6}) in the target space time. 

As our analysis reveals, this turns out to be a generic feature of the NR expansion (even for curved backgrounds) when the world-volume directions are considered to be stretched along longitudinal axes of the MNC manifold. Our natural expectation would therefore be to find some evidence for the nontrivial dynamics at NLO in the $ 1/c $ expansion. 
%%%%%%%%%%%%%%%%%%%%%%%%%%%%%%%%%%%%%%
\subsection{Dynamics at NLO}
The NLO Lagrangian density is identified as
\begin{eqnarray}
\tilde{\mathcal{L}}^{(-1)}=\frac{1}{2}\sqrt{-\det \tau_{mn}^{(2)}}\tilde{\tau}^{mn(2)}H_{mn}^{(-1)}+h^{(0)}.
\end{eqnarray}

After some simplification, one finds
\begin{eqnarray}
\label{3.14}
\tilde{\mathcal{L}}^{(-1)}=-\frac{x'^u x'^v}{2 \dot{x}^t}(\dot{x}^I)^2 +\frac{\dot{x}^t}{2} \left( \frac{x'^v}{x'^u}(\partial_1x^I)^2 + \frac{x'^u}{x'^v}(\partial_2 x^I)^2 \right) +\hat{C}\dot{x}^t x'^u x'^v,
\end{eqnarray}
where we choose $ \hat{C}_{tuv}=\hat{C}= $ constant.

Clearly, using the static gauge, one can imagine a simplest embedding of the form
\begin{eqnarray}
\label{3.15}
x^t = \sigma^0 ~;~ x^u = \sigma^1 ~;~ x^v = \sigma^2 ,
\end{eqnarray}
which can be further used to simplify (\ref{3.14}) to yield
\begin{eqnarray}
\label{3.16}
\tilde{\mathcal{L}}^{(-1)}=\frac{1}{2} \gamma^{mn}\partial_m x^I \partial_n x^I + \hat{C},
\end{eqnarray}
where $ \gamma^{mn}= \text{diag}(-1,1,1) $ is the conformally flat metric on the world-volume.

The above equation (\ref{3.16}), is pretty much in the Polyakov form (expressed using the conformal gauge) where the background three form has components only along the longitudinal directions of the MNC target space. The Lagrangian (\ref{3.16}) dictates the leading order dynamics of the transverse fluctuations associated with the NR M2 brane
\begin{eqnarray}
\partial_m (\gamma^{mn}\partial_n x^I)=0.
\end{eqnarray}

The Lagrangian (\ref{3.16}) enjoys a $ SO(1,2)\times SO(8) $ rotational symmetry. While $ SO(1,2) $ acts on the world-volume indices ($ m ,n =0,1,2$), the $ SO(8) $ acts on eight transverse directions ($ I=3, \cdots , 10 $). On top of this, the theory also posses a translation invariance, $ \delta x^I =a^I $ associated with the transverse directions of the brane.
%%%%%%%%%%%%%%%%%%%%%%%%%%%%%%%%%%%%%%
\subsection{Dynamics at NNLO}
The NNLO Lagrangian density could be formally expressed as
\begin{eqnarray}
\tilde{\mathcal{L}}^{(-2)}=\frac{1}{2}\sqrt{-\det \tau_{mn}^{(2)}}\tilde{\tau}^{mn(2)}\tau_{mn}^{(-2)}+h^{(-1)}.
\end{eqnarray}

After some simplification and using the ansatz (\ref{3.15}), one finds
\begin{eqnarray}
\label{3.19}
\tilde{\mathcal{L}}^{(-2)}=\frac{1}{2}\left( \gamma^{mn}\partial_m y^{\hat{A}}\partial_n y^{\hat{B}}\eta_{\hat{A}\hat{B}}-\epsilon^{mnp}\epsilon_{\hat{A}\hat{B}\hat{C}}\partial_mx^{\hat{A}}\partial_n y^{\hat{B}}\partial_p y^{\hat{C}}\right), 
\end{eqnarray}
where, we introduce the longitudinal metric $ \eta_{\hat{A}\hat{B}}= \text{diag}(-1,1,1)$ together with the longitudinal indices $ \hat{A},\hat{B}=t,u,v \equiv t, a$.

Therefore, combining (\ref{3.12}), (\ref{3.16}) and (\ref{3.19}) together one could recast the NR expansion (\ref{2.26}) as
\begin{eqnarray}
\mathcal{L}_{M2}=\left( \frac{1}{2} \gamma^{mn}\partial_m x^I \partial_n x^I + \hat{C}\right)~~~~~~~~~~~~~~~~~~~~~~~~~~~~~~~~~~~~~~~~~~~~~\nonumber\\
 +\frac{1}{2c}\left( \gamma^{mn}\partial_m y^{\hat{A}}\partial_n y^{\hat{B}}\eta_{\hat{A}\hat{B}}-\epsilon^{mnp}\epsilon_{\hat{A}\hat{B}\hat{C}}\partial_mx^{\hat{A}}\partial_n y^{\hat{B}}\partial_p y^{\hat{C}}\right)+\mathcal{O}(c^{-2}).
\end{eqnarray}

Like in the previous example, the NNLO Lagrangian (\ref{3.19}) is also invariant under $ SO(1,2) $ and $ SO(8) $ global rotations. The dynamics of the corresponding world-volume fields is governed by the following equation
\begin{eqnarray}
\partial_m (\gamma^{mn}\partial_n y^{\hat{B}}\eta_{\hat{A}\hat{B}})-\partial_m (\epsilon^{mnp}\epsilon_{\hat{A}\hat{B}\hat{C}}\partial_n x^{\hat{B}}\partial_p y^{\hat{C}})=0.
\end{eqnarray}
Clearly, the NNLO theory governs the dynamics of longitudinal fluctuations ($ y^{\hat{A}} $) along the world-volume directions of the NR M2 brane. 
%%%%%%%%%%%%%%%%%%%%%%%%%%%%%%%%%%%%%%
\section{Nonrelativistic limit of M2 branes in $ AdS_4 \times S^7 $}
\label{sec4}
Having explored the flat space dynamics, we now move on towards the most generic situation namely the M2 brane dynamics on curved backgrounds. The strategy would be to start with the relativistic M2 brane solutions in $ AdS_4 \times S^7 $ backgrounds and thereby taking a NR limit of the background solution following the algorithm of \cite{Blair:2021waq},\cite{Hartong:2021ekg}.

The background can be expressed in global coordinates as \cite{Bozhilov:2005ew}
\begin{eqnarray}
ds^2_{11}&=&L^2 (-dt^2 \cosh^2 \rho +d \rho^2 + \sinh^2\rho (d\alpha^2 + \sin^2\alpha d\beta^2) +ds^2_7),\\
C_3 &=& -\frac{k}{3}\sinh^3\rho \sin\alpha dt \wedge d\alpha \wedge d\beta .
\end{eqnarray}
%%%%%%%%%%%%%%%%%%%%%%%%%%%%%%%%%%
\subsection{MNC data}
We first decode the longitudinal data in the NR limit where we assume that the M2 brane directions are along the directions of $ AdS_4 $. The seven sphere $ S^7 $ is considered to be the space transverse to the world-volume directions of the M2 brane. However, in the present analysis, we restrict the dynamics of the M2 brane only in the $ AdS_4 $ where one of the world-volume direction of the M2 brane wraps the isometry direction $ \beta (\in S^2 )$. 

We consider the following NR scaling of the the target space directions\footnote{ Clearly, the $ c^{-2} $ scaling of $ \alpha $ in (\ref{4.4}) produces a $ \mathcal{O}(c^{-4}) $ correction in the expansion for the three form flux field (\ref{4.17}) which is therefore disregarded here.}
\begin{eqnarray}
L &=& c \ell ~;~ t= t^{(0)}+\frac{1}{c^2}t^{(1)}~;~\rho = \rho^{(0)}+\frac{1}{c^2}\rho^{(1)},\\
\alpha &=& \alpha^{(0)} +\frac{1}{c^{2}} \alpha^{(1)}~;~\beta = \beta^{(0)}+\frac{1}{c^2}\beta^{(1)},
\label{4.4}
\end{eqnarray}
where we set $ \alpha^{(0)} = \frac{\pi}{2}=$ constant and as per our previous notation, $ \rho =u $ and $ \beta = v $.

Below we note down the longitudinal components of the metric
\begin{eqnarray}
\label{4.5}
\tau_{tt}=  -\ell^2 \cosh ^2\rho^{(0)}~;~\tau_{\rho \rho}=\ell^2 ~;~ \tau_{\beta \beta}=\ell^2 \sinh^2\rho^{(0)}.
\end{eqnarray}

Using (\ref{4.5}), one obtains the following MNC data
\begin{eqnarray}
\tau_{t}~^0 = \ell \cosh \rho^{(0)}~;~\tau_{\rho}~^{1}= \ell ~;~ \tau_{\beta}~^{2}= \ell \sinh \rho^{(0)}.
\end{eqnarray}

Finally, we note down the longitudinal metric along the world-volume directions of the NR M2 brane\footnote{Here, we assume that all the leading order transverse fluctuations ($ x^I $) are freezed out and the NR M2 brane has fluctuations only along the longitudinal directions ($ x^A , y^A $).}
\begin{eqnarray}
\label{4.7}
\tau_{00}^{(2)}&=&\kappa^2 \tau_{tt}~;~\tau_{11}^{(2)}=\tau_{\rho \rho}(\partial_1 \rho^{(0)})^2~;~\tau_{22}^{(2)}=\tau_{\beta \beta}(\partial_2 \beta^{(0)})^2~;~H_{mn}^{(-1)}=0,\\
\tau_{00}^{(-2)}&=& \tau_{tt}~;~\tau_{ij}^{(-2)}=\tau_{ab}\partial_i y^a \partial_j y^b~;~x^a = \lbrace \rho^{(0)}, \beta^{(0)} \rbrace~;~y^a = \lbrace \rho^{(1)}, \beta^{(1)} \rbrace ,\\
\tau_{00}^{(0)}&=&2 \kappa \tau_{tt}~;~\tau_{0i}^{(0)}=0~;~\tau_{ij}^{(0)}=\tau_{ab}(\partial_i x^a \partial_j y^b +\partial_i y^a \partial_j x^b),
\label{4.9}
\end{eqnarray}
which is subjected to the realization
\begin{eqnarray}
\label{4.10}
x^t &=& t^{(0)}=\kappa \sigma^{0}~;~y^t = t^{(1)}=\sigma^0 ~;~\rho^{(0)} =\rho^{(0)}(\sigma^1) ,\\
\rho^{(1)} &=&\rho^{(1)}(\sigma^1)~;~\beta^{(0)} = b^{(0)}\sigma^2~;~\beta^{(1)}= b^{(1)}\sigma^{2},
\label{4.11}
\end{eqnarray}
where $ b^{(0)} $ and $ b^{(1)} $ are the respective winding numbers.

Here, one of the world-volume directions $ \sigma^1 $ is extended along the AdS radial direction ($ \rho $) while the other compact direction is wrapping the isometry of $ S^2 $. However, in our analysis, we consider folded NR M2 brane solutions which is subjected to the periodicity condition along the radial direction 
\begin{eqnarray}
\rho (\sigma^1 + 2\pi)=\rho (\sigma^1).
\end{eqnarray}
%%%%%%%%%%%%%%%%%%%%%%%%%%%%%%%%%%
\subsection{Dynamics at LO}
Given the above set of data (\ref{4.7})-(\ref{4.9}), we note down the following
\begin{eqnarray}
\frac{1}{2}\sqrt{-\det \tau_{mn}^{(2)}}\tilde{\tau}^{mn(2)}\tau_{mn}^{(0)}&=&\frac{\kappa b^{(0)}}{2}\rho'^{(0)}\sinh 2\rho^{(0)} \left(\frac{1}{\kappa}+\frac{\rho'^{(1)}}{\rho'^{(0)}}+\frac{b^{(1)}}{b^{(0)}} \right), \\
h^{(1)}&=&-\frac{\kappa b^{(0)}}{2}\rho'^{(0)}\sinh 2\rho^{(0)} \left(\frac{1}{\kappa}+\frac{\rho'^{(1)}}{\rho'^{(0)}}+\frac{b^{(1)}}{b^{(0)}} \right),
\end{eqnarray}
where we set, $ \ell =1 $ for simplicity.

Clearly, when combined together, the zeroth order Lagrangian density vanishes
\begin{eqnarray}
\label{4.15}
\tilde{\mathcal{L}}^{(0)}=0.
\end{eqnarray}

The vanishing of the leading order Lagrangian (\ref{4.15}) is precisely due to the M2 brane embeddings considered in (\ref{4.10})-(\ref{4.11}). In other words, the membrane dynamics at leading order turns out to be trivial in the static gauge, where the world-volume directions are stretched along the longitudinal axes. Therefore, the evidence of non-trivial membrane dynamics appears to be only at NLO and beyond in the NR expansion.
%%%%%%%%%%%%%%%%%%%%%%%%%%%%%%%%%%
\subsection{Dynamics at NLO}
The NLO Lagrangian can be expressed as
\begin{eqnarray}
\tilde{\mathcal{L}}^{(-1)}=\frac{\epsilon^{mnp}}{6}\partial_m x^{\mu} \partial_n x^{\nu} \partial_p x^{\lambda} \hat{C}_{\mu \nu \lambda},
\end{eqnarray}
which is subjected to the fact, $ H_{mn}^{(-1)}=H_{IJ}\partial_m x^I \partial_n x^J =0 $. 

Background three form flux ($ C_3 $) has two of its legs along longitudinal directions $ (t , \beta) $ and the remaining one along the transverse axis $ \alpha $. The corresponding NR expansion yields
\begin{eqnarray}
\label{4.17}
C_{t\alpha \beta}=-\frac{k\alpha^{(1)}}{3c^4}\sinh^3 \rho^{(0)}+\cdots ,
\end{eqnarray}
where, $ \alpha^{(1)}$ is what we identify as the transverse fluctuations associated with the world-volume fields at NLO. Comparing with $ (\ref{2.8}) $, this further leads to, $ \hat{C}_{t\alpha \beta}=0 $.

Combining all these facts together, the NLO Lagrangian density turns out to be
\begin{eqnarray}
\tilde{\mathcal{L}}^{(-1)}=0.
\end{eqnarray}
%%%%%%%%%%%%%%%%%%%%%%%%%%%%%%%%%%
\subsection{Dynamics at NNLO}
The first nontrivial dynamics therefore appears to be at NNLO
\begin{eqnarray}
\label{4.19}
\tilde{\mathcal{L}}^{(-2)}=\frac{\sqrt{- \det \tau^{(2)}_{mn}}}{2}\tilde{\tau}^{mn(2)} \tau_{mn}^{(-2)} +h^{(-1)}.
\end{eqnarray}

A straightforward computation reveals the Lagrangian density of the form
\begin{eqnarray}
\label{4.20}
\tilde{\mathcal{L}}^{(-2)}&=&\frac{\kappa b^{(0)}}{4}\sinh 2\sigma^1 (\rho'^{2(1)}+\Gamma)-\frac{1}{2}\sinh 2\sigma^1 (g \rho'^{(1)}+b^{(1)}),\\
\Gamma &=& \frac{1}{\kappa^2}+\frac{b^{2(1)}}{b^{2(0)}} ~;~g=b^{(0)}+ \kappa b^{(1)},
\end{eqnarray}
where we set, $ \rho^{(0)} =\sigma^1$ and $ \frac{b^{(1)}}{b^{(0)}}=\frac{\Gamma \kappa}{2} $ without any loss of generality\footnote{One possible choice could be setting, $ \kappa = b^{(0)}=b^{(1)}=1 $.}.

The equation of motion that readily follows from (\ref{4.20}) could be expressed as
\begin{eqnarray}
\label{4.22}
\rho'^{(1)}-\frac{g}{\kappa b^{(0)}}=-\frac{c_1}{\sinh 2 \sigma^1 },
\end{eqnarray}
where $ c_1 $ being the constant of integration.

\paragraph{Folded M2 branes.} Notice that, the integral (\ref{4.22}) leads to a logarithmic divergence near $ \sigma^1 \sim \varepsilon \sim 0 $ which, as we shall see, can be overcome by fixing the intergation constant $ c_2 $. A straightforward integration of (\ref{4.22}) reveals
\begin{eqnarray}
\label{4.23}
\rho^{(1)}(\sigma^1)=\frac{g \sigma^1 }{\kappa b^{(0)} }+\frac{c_1}{2}  \log \coth \sigma^1 +c_2.
\end{eqnarray}

By demanding the fact that $ \rho^{(1)}(\sigma^1)|_{\sigma^1 \sim \varepsilon}=0 $, we fix the one of the integration constants 
\begin{eqnarray}
c_2 = \frac{c_1}{2}\log \varepsilon .
\end{eqnarray}

On the other hand, setting 
\begin{eqnarray}
\rho'(\sigma^1)|_{\sigma^1 = \pi/2}=1+c^{-2} \rho'^{(1)}|_{\sigma^1 = \pi/2} =0,
\end{eqnarray}
one can fix the other integration constant as
\begin{eqnarray}
c_1= \left(c^2+\frac{g}{\kappa b^{(0)}  }\right)\sinh\pi .
\end{eqnarray}

Therefore, the complete radial solution for these closed NR M2 brane configuration is given by\footnote{The logarithmic divergence might well be an artefact of some geometric singularity near the origin $ \rho \sim 0 $ of the bulk spacetime. This can be seen from the structure of the inverse vielbeins for example in (\ref{4.37}). However, a proper understanding of the nature of this singularity seeks further investigations.}
\begin{eqnarray}
\rho (\sigma^1)=\left(  \sigma^1 +\frac{1}{2}\sinh\pi  \log \coth \sigma^1 +\frac{1}{2}\log \varepsilon \right) ~~~~~~\nonumber\\
+ c^{-2}\left(\frac{g \sigma^1 }{\kappa b^{(0)} }+\frac{g}{2\kappa b^{(0)}  }\sinh\pi  (\log \coth \sigma^1 +\log\varepsilon)\right).
\end{eqnarray}

\paragraph{Spectrum.} The energy associated with the folded M2 branes is given by
\begin{eqnarray}
\mathcal{E}_{NR}=\frac{1}{8} \tanh \sigma^1 (\cosh \sigma^1 (\sinh \pi  (\log \coth \sigma^1 +\log \varepsilon)-2)-\sigma^1  \sinh \pi  \text{csch}\sigma^1 )\\
\times (\cosh \sigma^1 (\sinh \pi  (\log \coth \sigma^1 +\log \varepsilon )+2)-\sigma^1  \sinh \pi \text{csch}\sigma^1 ).
\end{eqnarray}
where we set $ \kappa = b^{(0)}=b^{(1)}=1 $ for simplicity.

The total energy of the configuration is obtained by integrating over the compact world-volume directions
\begin{eqnarray}
\label{4.30}
E_{NR}=\frac{\tilde{T}_2}{2\pi c^2}\int d\sigma^2 d\sigma^1 \mathcal{E}_{NR}=\frac{\tilde{T}_2}{c^2}\int_0^{2\pi}  d\sigma^1 \mathcal{E}_{NR}.
\end{eqnarray}

The integral (\ref{4.30}) can be evaluated for short membranes which yields
\begin{eqnarray}
E_{NR}=\frac{\pi \tilde{T}_2}{8c^2}\left(\mathcal{C} \sinh^2 \pi  -8 + \cdots \right), 
\end{eqnarray}
where $ \mathcal{C} $ is a constant.
%%%%%%%%%%%%%%%%%%%%%%%%%%%%%%%%%%
\section{Nonrelativistic spinning membranes}
\label{sec5}
We now generalise the NR folded M2 brane solutions in the presence of a spin ($ S $) which is turned around the compact isometry direction ($ \beta $) of the MNC manifold. To this end, we choose to work with an ansatz of the form
\begin{eqnarray}
 t^{(0)}=\kappa \sigma^{0}~;~ t^{(1)}=\sigma^0 ~;~\rho^{(0)} =\rho^{(0)}(\sigma^1) ~;~\rho^{(1)} =\rho^{(1)}(\sigma^1),\\
\beta^{(0)}(\sigma^0 , \sigma^2) =\omega \sigma^0 + b^{(0)}\sigma^2~;~\beta^{(1)}(\sigma^0 , \sigma^2) = \hat{\omega} \sigma^0 +b^{(1)}\sigma^{2},
\end{eqnarray}
where $ \omega $ is the angular frequency along the compact isometry $ \beta $.
%%%%%%%%%%%%%%%%%%%%%%%%%%%%%%%%%%%%%%%%%%%%%
\subsection{Background data}
In order to proceed further, we first list down the NR background data 
\begin{eqnarray}
\label{4.34}
\tau_{00}^{(2)}&=&\kappa^2 \tau_{tt}+\omega^2 \tau_{\beta \beta}~;~\tau_{11}^{(2)}=\tau_{\rho \rho}(\partial_1 \rho^{(0)})^2~;~\tau_{22}^{(2)}=\tau_{\beta \beta}~;~\tau_{02}^{(2)}=\omega \tau_{\beta \beta},\\
\tau_{00}^{(-2)}&=& \tau_{tt}+ \hat{\omega}^2 \tau_{\beta \beta}~;~\tau_{ij}^{(-2)}=\tau_{ab}\partial_i y^a \partial_j y^b~;~\tau_{02}^{(-2)}= \hat{\omega} \tau_{\beta \beta}~;~H_{mn}^{(-1)}=0,\\
\tau_{00}^{(0)}&=&2 \kappa \tau_{tt}+ 2 \omega \hat{\omega} \tau_{\beta \beta}~;~\tau_{02}^{(0)}=(\omega + \hat{\omega}) \tau_{\beta \beta}~;~\tau_{ij}^{(0)}=\tau_{ab}(\partial_i x^a \partial_j y^b +\partial_i y^a \partial_j x^b),
\label{4.36}
\end{eqnarray}
where for simplicity we set, $ b^{(0)}=b^{(1)}=1 $ without any loss of generality.

Using the data (\ref{4.34}), one can further obtain the inverse $ 3 \times 3 $ matrix as
\begin{eqnarray}
\label{4.37}
\tilde{\tau}^{mn(2)}_{3 \times 3}=\left(
\begin{array}{ccc}
 \frac{1}{\kappa ^2 \tau _{\text{tt}}} & 0 & -\frac{\omega }{\kappa ^2 \tau _{\text{tt}}} \\
 0 & \frac{1}{\left(\partial_1 \rho^{(0)}\right)^2 \tau _{\rho \rho }} & 0 \\
 -\frac{\omega }{\kappa ^2 \tau _{\text{tt}}} & 0 & \frac{\omega ^2}{\kappa ^2 \tau _{\text{tt}}}+\frac{1}{\tau _{\beta \beta }} \\
\end{array}
\right).
\end{eqnarray}

Given the background data (\ref{4.34})-(\ref{4.37}), our next task would be to obtain the Lagrangian densities at different order in the $ 1/c $ expansion.
%%%%%%%%%%%%%%%%%%%%%%%%%%%%%%%%%%%%%%%%%%%%%%
\subsection{Dynamics at LO} 
Like before, we note down the following
\begin{eqnarray}
\label{5.7}
\frac{1}{2}\sqrt{-\det \tau_{mn}^{(2)}}\tilde{\tau}^{mn(2)}\tau_{mn}^{(0)}
&=&\frac{1}{2}\sinh 2 \rho^{(0)}(\kappa (\rho'^{(1)}+ \rho'^{(0)})+\rho'^{(0)}),\\ 
\label{5.8}
h^{(1)}&=&-\frac{1}{2}\sinh 2 \rho^{(0)}(\kappa (\rho'^{(1)}+ \rho'^{(0)})+\rho'^{(0)}).
\end{eqnarray}

Combining (\ref{5.7})-(\ref{5.8}), the leading order Lagrangian density vanishes
\begin{eqnarray}
\label{5.9}
\tilde{\mathcal{L}}^{(0)}=0,
\end{eqnarray}
which is identical to the example we have studied in the previous section.

As before, if we proceed further, the NLO Lagrangian density vanishes identically, $\tilde{\mathcal{L}}^{(-1)}= 0 $ as the leading order transverse fluctuations are set to zero. Therefore, the next non-trivial correction to the world-volume theory comes at NNLO which we study next.
%%%%%%%%%%%%%%%%%%%%%%%%%%%%%%%%%%%%%%%%%%%%%%
\subsection{Dynamics at NNLO} 
The nontrivial dynamics appears to be at NNLO that is accompanied by a Lagrangian density of the form (\ref{4.19}). Below, we enumerate each of the terms individually
\begin{eqnarray}
\frac{\sqrt{- \det \tau^{(2)}_{mn}}}{2}\tilde{\tau}^{mn(2)} \tau_{mn}^{(-2)} &=&\frac{\kappa}{4}\sinh 2\sigma^1 \left((\rho'^{(1)})^2 +1 + \frac{1}{\kappa^2}\left(1-(\omega - \hat{\omega})^2 \tanh^2\sigma^1 \right)  \right) ,\\
h^{(-1)}&=&-\frac{1}{2}\sinh 2\sigma^1 (\rho'^{(1)}(\kappa +1)+1),
\end{eqnarray}
where, we set $ \rho^{(0)}=\sigma^1 $ without any loss of generality.

The equation of motion corresponding to $ \rho^{(1)} $ yields a simple form
\begin{eqnarray}
\rho'^{(1)}(\sigma^1)= \frac{c_1}{\sinh 2\sigma^1}+\frac{(\kappa +1)}{\kappa},
\end{eqnarray}
which exhibits a solution similar to that of (\ref{4.23}).
%%%%%%%%%%%%%%%%%%%%%%%%%%%%%%%%%%%%%%
\subsection{Dispersion relation}
The energy and the spin of the NR M2 brane configuration is given by the following definitions (where we scale the entities by an overall factor of $ \frac{\tilde{T}_2}{c^2} $)
\begin{eqnarray}
\label{5.13}
E_{NR}&=&\frac{1}{2\pi}\int d\sigma^2 d\sigma^1 \mathcal{E}_{NR}=\int_0^{2\pi}  d\sigma^1 \mathcal{E}_{NR},\\
S&=&\frac{1}{2\pi }\int d\sigma^2 d\sigma^1 \mathcal{S}=\int_0^{2\pi}  d\sigma^1 \mathcal{S}. 
\label{5.14}
\end{eqnarray}

The densities can be expressed as
\begin{eqnarray}
\label{5.15}
\mathcal{E}_{NR}=\frac{\delta \tilde{\mathcal{L}}^{(-2)}}{\delta \dot{t}^{(0)}}=\frac{1}{4}(\rho'^{2(1)}+1)\sinh 2\sigma^1 - \frac{\rho'^{(1)}}{2}\sinh 2\sigma^1 \nonumber\\
- \frac{1}{4 \kappa^2}\sinh 2\sigma^1\left(1-(\omega - \hat{\omega})^2 \tanh^2\sigma^1 \right),
\end{eqnarray}
\begin{eqnarray}
\label{5.16}
\mathcal{S}=\frac{\delta \tilde{\mathcal{L}}^{(-2)}}{\delta \dot{\beta}^{(0)}}=-\frac{(\omega - \hat{\omega})}{2\kappa}\tanh^2 \sigma^1 \sinh 2\sigma^1.
\end{eqnarray}

Plugging (\ref{5.15})-(\ref{5.16}) into their respective integrals (\ref{5.13}) and (\ref{5.14}), we find
\begin{eqnarray}
\label{5.17}
E_{NR}&=& \frac{(\omega - \hat{\omega})^2}{8 \kappa^2}(-1+\cosh 4 \pi +4 \log (\text{sech}2 \pi ))+\cdots ,\\
S&=&-\frac{(\omega - \hat{\omega})}{4 \kappa}(-1+\cosh 4 \pi +4 \log (\text{sech}2 \pi )),
\label{5.18}
\end{eqnarray}
where we ignore remaining terms in (\ref{5.17}) being constants only.

Combining (\ref{5.17}) and (\ref{5.18}), we finally obtain the NR dispersion relation 
\begin{eqnarray}
E_{NR} \approx \gamma_{0} S^2 ,
\end{eqnarray}
where the constant $\gamma_0  $ could be easily read off from the above formulae (\ref{5.17}) and (\ref{5.18}).
%%%%%%%%%%%%%%%%%%%%%%%%%%%%%%%%%%%%%%
\section{Concluding remarks and future directions}
\label{sec6}
We conclude by stating the main results of this paper. The present paper investigates the dynamics of nonrelativistic (NR) M2 brane solutions those propagate over M2 brane Newton Cartan (MNC) backgrounds. These backgrounds are obtained using the NR scaling of the background fields of an eleven dimensional M theory background. Finally, the NR world-volume theory is obtained by taking into account a $ 1/c^2 $ expansion \cite{Hartong:2021ekg} for the embedding fields those are living on M2 branes propagating over MNC target space.

We choose to work with a particular embedding and show that the leading order (LO) action vanishes (in the static gauge) when the world-volume axes of the M2 brane are extended along the longitudinal axes. We confirm this claim by constructing the world-volume theory both for the flat as well as the curved MNC manifold.

However, as we show schematically below, this is a generic feature of NR M2 branes for any MNC background. A straightforward calculation using longitudinal vielbeins reveals 
\begin{align}
\label{6.1}
\frac{1}{2}\det (\tau^{(2)})_m^A \tilde{\tau}^{(2)mn}\tau_{mn}^{(0)}&=\frac{1}{2}\varepsilon^{abc}\Big[ \tau_a~^{(2)0}(\tau_b~^{(2)1}\tau_c~^{(2)2}-\tau_b~^{(2)2}\tau_c~^{(2)1})\nonumber\\&+\cdots \Big]\tilde{\tau}^{(2)mn}\tau_{mn}^{(0)}\nonumber\\
&=-\frac{1}{2}\varepsilon^{abc}\tau_b~^{(2)1}\tau_c~^{(2)2}\tau^{n(2)}~_0 \tau_{an}^{(0)}+\cdots \nonumber\\
&=\frac{1}{2}\varepsilon^{abc}\tau_b~^{(2)1}\tau_c~^{(2)2}\tau_\nu~^0 \partial_a y^\nu + \cdots ,
\end{align}
where we use the notation $ \tau_a~^{(2)A}=\tau_\mu~^A \partial_a x^\mu $ together with the fact
\begin{align}
\tau_{an}^{(0)}=\tau_a~^{(2)A}\tau_\nu~^B \eta_{AB}\partial_n y^\nu.
\end{align}

On the other hand, a parallel computation reveals
\begin{align}
\label{6.3}
h^{(1)}&=-\frac{1}{2}\varepsilon^{mnp}\tau_\mu~^1 (\tau_\nu~^2 \tau_\lambda~^0 -\tau_\nu~^0 \tau_\lambda~^2)\partial_p y^\lambda \partial_m x^\mu \partial_n x^\nu +\cdots \nonumber\\
&=-\frac{1}{2}\varepsilon^{pmn}\tau_m~^{(2)1}\tau_n~^{(2)2}\tau_\lambda~^0 \partial_p y^\lambda + \cdots .
\end{align}
Combining, (\ref{6.1}) and (\ref{6.3}) one might therefore argue that the leading order Lagrangian density ($ \tilde{\mathcal{L}}^{(0)}  $) vanishes identically for generic MNC backgrounds (and without choosing any flat gauge) which clearly agrees with the NR expansion for M2 branes in flat space \cite{Garcia:2002fa}.

There are some possible extensions of the present work those might be of worth exploring in the near future. One of them is certainly the consideration of the dimensional reductions along the longitudinal as well as the transverse axes of the MNC manifold. 

A natural expectation would to find an analogue of String Newton-Cartan (SNC) limit while reducing the theory along one of the longitudinal axes. On the other hand, dimensional reduction along one of the transverse directions should lead to a NR world-volume theory whose dynamics should be reproducible by considering the NR limit of a single D2 brane (in ten dimensions) in a $ 1/c $ expansion. 

In case of longitudinal reduction, the resulting expansion should agree to those obtained in \cite{Hartong:2021ekg}. On the other hand, for a single D2 brane, a $ 1/c $ expansion is yet to be constructed.

Finally, it would be nice to find an analogue of the decompactification or the large $ R $ limit \cite{Hartong:2021ekg} for M2 branes. This analogy might well be the case as the M2 branes over a noncompact (flat) target space and in the presence of a constant three form flux shows the existence of string like spikes in its spectrum \cite{GarciaDelMoral:2018jye}. This seeks a deeper investigation of the M2 branes and should be an interesting direction to look for in the future. 
%%%%%%%%%%%%%%%%%%%%%%%%%%%
\acknowledgments
The author is indebted to the authorities of IIT Roorkee for their unconditional support towards researches in basic sciences. The author acknowledges The Royal Society, UK for financial assistance and the Grant (No. SRG/2020/000088) received from The Science and Engineering Research Board (SERB), India.
%%%%%%%%%%%%%%%%%%%%%%%%%%%%%%


\begin{thebibliography}{99}
\bibitem{Gomis:2000bd}
J.~Gomis and H.~Ooguri,
``Nonrelativistic closed string theory,''
J. Math. Phys. \textbf{42}, 3127-3151 (2001)
doi:10.1063/1.1372697
[arXiv:hep-th/0009181 [hep-th]].

\bibitem{Danielsson:2000gi}
U.~H.~Danielsson, A.~Guijosa and M.~Kruczenski,
``IIA/B, wound and wrapped,''
JHEP \textbf{10}, 020 (2000)
doi:10.1088/1126-6708/2000/10/020
[arXiv:hep-th/0009182 [hep-th]].

\bibitem{Gomis:2005pg}
J.~Gomis, J.~Gomis and K.~Kamimura,
``Non-relativistic superstrings: A New soluble sector of AdS(5) x S**5,''
JHEP \textbf{12}, 024 (2005)
doi:10.1088/1126-6708/2005/12/024
[arXiv:hep-th/0507036 [hep-th]].

\bibitem{Bergshoeff:2015uaa}
E.~Bergshoeff, J.~Rosseel and T.~Zojer,
``Newton\textendash{}Cartan (super)gravity as a non-relativistic limit,''
Class. Quant. Grav. \textbf{32}, no.20, 205003 (2015)
doi:10.1088/0264-9381/32/20/205003
[arXiv:1505.02095 [hep-th]].

\bibitem{Bergshoeff:2021bmc}
E.~A.~Bergshoeff, J.~Lahnsteiner, L.~Romano, J.~Rosseel and C.~\c{S}im\c{s}ek,
``A non-relativistic limit of NS-NS gravity,''
JHEP \textbf{06}, 021 (2021)
doi:10.1007/JHEP06(2021)021
[arXiv:2102.06974 [hep-th]].

\bibitem{Hansen:2019pkl}
D.~Hansen, J.~Hartong and N.~A.~Obers,
``Action Principle for Newtonian Gravity,''
Phys. Rev. Lett. \textbf{122}, no.6, 061106 (2019)
doi:10.1103/PhysRevLett.122.061106
[arXiv:1807.04765 [hep-th]].

\bibitem{Bergshoeff:2019pij}
E.~A.~Bergshoeff, J.~Gomis, J.~Rosseel, C.~Simsek and Z.~Yan,
``String Theory and String Newton-Cartan Geometry,''
J. Phys. A \textbf{53}, no.1, 014001 (2020)
doi:10.1088/1751-8121/ab56e9
[arXiv:1907.10668 [hep-th]].

\bibitem{Bergshoeff:2018yvt}
E.~Bergshoeff, J.~Gomis and Z.~Yan,
``Nonrelativistic String Theory and T-Duality,''
JHEP \textbf{11}, 133 (2018)
doi:10.1007/JHEP11(2018)133
[arXiv:1806.06071 [hep-th]].

\bibitem{Harmark:2017rpg}
T.~Harmark, J.~Hartong and N.~A.~Obers,
``Nonrelativistic strings and limits of the AdS/CFT correspondence,''
Phys. Rev. D \textbf{96}, no.8, 086019 (2017)
doi:10.1103/PhysRevD.96.086019
[arXiv:1705.03535 [hep-th]].

\bibitem{Harmark:2018cdl}
T.~Harmark, J.~Hartong, L.~Menculini, N.~A.~Obers and Z.~Yan,
``Strings with Non-Relativistic Conformal Symmetry and Limits of the AdS/CFT Correspondence,''
JHEP \textbf{11}, 190 (2018)
doi:10.1007/JHEP11(2018)190
[arXiv:1810.05560 [hep-th]].

\bibitem{Roychowdhury:2019sfo}
D.~Roychowdhury,
``Semiclassical dynamics for torsional Newton-Cartan strings,''
Nucl. Phys. B \textbf{958}, 115132 (2020)
doi:10.1016/j.nuclphysb.2020.115132
[arXiv:1911.10473 [hep-th]].

\bibitem{Roychowdhury:2019olt}
D.~Roychowdhury,
``Nonrelativistic pulsating strings,''
JHEP \textbf{09}, 002 (2019)
doi:10.1007/JHEP09(2019)002
[arXiv:1907.00584 [hep-th]].

\bibitem{Yan:2021hte}
Z.~Yan and M.~Yu,
``KLT factorization of nonrelativistic string amplitudes,''
JHEP \textbf{04}, 068 (2022)
doi:10.1007/JHEP04(2022)068
[arXiv:2112.00025 [hep-th]].

\bibitem{Gomis:2020fui}
J.~Gomis, Z.~Yan and M.~Yu,
``Nonrelativistic Open String and Yang-Mills Theory,''
JHEP \textbf{03}, 269 (2021)
doi:10.1007/JHEP03(2021)269
[arXiv:2007.01886 [hep-th]].

\bibitem{Gomis:2020izd}
J.~Gomis, Z.~Yan and M.~Yu,
``T-Duality in Nonrelativistic Open String Theory,''
JHEP \textbf{02}, 087 (2021)
doi:10.1007/JHEP02(2021)087
[arXiv:2008.05493 [hep-th]].

\bibitem{Hartong:2021ekg}
J.~Hartong and E.~Have,
``Nonrelativistic Expansion of Closed Bosonic Strings,''
Phys. Rev. Lett. \textbf{128}, no.2, 021602 (2022)
doi:10.1103/PhysRevLett.128.021602
[arXiv:2107.00023 [hep-th]].

\bibitem{Harmark:2019upf}
T.~Harmark, J.~Hartong, L.~Menculini, N.~A.~Obers and G.~Oling,
``Relating non-relativistic string theories,''
JHEP \textbf{11}, 071 (2019)
doi:10.1007/JHEP11(2019)071
[arXiv:1907.01663 [hep-th]].

\bibitem{Gallegos:2019icg}
A.~D.~Gallegos, U.~G\"ursoy and N.~Zinnato,
``Torsional Newton Cartan gravity from non-relativistic strings,''
JHEP \textbf{09}, 172 (2020)
doi:10.1007/JHEP09(2020)172
[arXiv:1906.01607 [hep-th]].

\bibitem{Gomis:2019zyu}
J.~Gomis, J.~Oh and Z.~Yan,
``Nonrelativistic String Theory in Background Fields,''
JHEP \textbf{10}, 101 (2019)
doi:10.1007/JHEP10(2019)101
[arXiv:1905.07315 [hep-th]].

\bibitem{Bidussi:2021ujm}
L.~Bidussi, T.~Harmark, J.~Hartong, N.~A.~Obers and G.~Oling,
``Torsional string Newton-Cartan geometry for non-relativistic strings,''
JHEP \textbf{02}, 116 (2022)
doi:10.1007/JHEP02(2022)116
[arXiv:2107.00642 [hep-th]].

\bibitem{Yan:2021lbe}
Z.~Yan,
``Torsional deformation of nonrelativistic string theory,''
JHEP \textbf{09}, 035 (2021)
doi:10.1007/JHEP09(2021)035
[arXiv:2106.10021 [hep-th]].

\bibitem{Roychowdhury:2019qmp}
D.~Roychowdhury,
``Probing tachyon kinks in Newton-Cartan background,''
Phys. Lett. B \textbf{795}, 225-229 (2019)
doi:10.1016/j.physletb.2019.06.031
[arXiv:1903.05890 [hep-th]].

\bibitem{Kamimura:2005rz}
K.~Kamimura and T.~Ramirez,
``Brane dualities in non-relativistic limit,''
JHEP \textbf{03}, 058 (2006)
doi:10.1088/1126-6708/2006/03/058
[arXiv:hep-th/0512146 [hep-th]].

\bibitem{Ebert:2021mfu}
S.~Ebert, H.~Y.~Sun and Z.~Yan,
``Dual D-brane actions in nonrelativistic string theory,''
JHEP \textbf{04}, 161 (2022)
doi:10.1007/JHEP04(2022)161
[arXiv:2112.09316 [hep-th]].

\bibitem{Blair:2021waq}
C.~D.~A.~Blair, D.~Gallegos and N.~Zinnato,
``A non-relativistic limit of M-theory and 11-dimensional membrane Newton-Cartan geometry,''
JHEP \textbf{10}, 015 (2021)
doi:10.1007/JHEP10(2021)015
[arXiv:2104.07579 [hep-th]].

\bibitem{Bergshoeff:2022pzk}
E.~Bergshoeff, J.~Lahnsteiner, L.~Romano and J.~Rosseel,
``The Supersymmetric Neveu-Schwarz Branes of Non-Relativistic String Theory,''
[arXiv:2204.04089 [hep-th]].

\bibitem{Sezgin:2002rt}
E.~Sezgin and P.~Sundell,
``Massless higher spins and holography,''
Nucl. Phys. B \textbf{644}, 303-370 (2002)
[erratum: Nucl. Phys. B \textbf{660}, 403-403 (2003)]
doi:10.1016/S0550-3213(02)00739-3
[arXiv:hep-th/0205131 [hep-th]].

\bibitem{Brugues:2004pj}
J.~Brugues, J.~Rojo and J.~G.~Russo,
``Non-perturbative states in type II superstring theory from classical spinning membranes,''
Nucl. Phys. B \textbf{710}, 117-138 (2005)
doi:10.1016/j.nuclphysb.2005.01.019
[arXiv:hep-th/0408174 [hep-th]].

\bibitem{Alishahiha:2002sy}
M.~Alishahiha and M.~Ghasemkhani,
``Orbiting membranes in M theory on AdS(7) x S**4 background,''
JHEP \textbf{08}, 046 (2002)
doi:10.1088/1126-6708/2002/08/046
[arXiv:hep-th/0206237 [hep-th]].

\bibitem{Alishahiha:2002fi}
M.~Alishahiha and A.~E.~Mosaffa,
``Circular semiclassical string solutions on confining AdS / CFT backgrounds,''
JHEP \textbf{10}, 060 (2002)
doi:10.1088/1126-6708/2002/10/060
[arXiv:hep-th/0210122 [hep-th]].

\bibitem{Bozhilov:2003wr}
P.~Bozhilov,
``M2-brane solutions in AdS(7) x S**4,''
JHEP \textbf{10}, 032 (2003)
doi:10.1088/1126-6708/2003/10/032
[arXiv:hep-th/0309215 [hep-th]].

\bibitem{Bozhilov:2005ew}
P.~Bozhilov,
``Membrane solutions in M-theory,''
JHEP \textbf{08}, 087 (2005)
doi:10.1088/1126-6708/2005/08/087
[arXiv:hep-th/0507149 [hep-th]].

\bibitem{Garcia:2002fa}
J.~A.~Garcia, A.~Guijosa and J.~D.~Vergara,
``A Membrane action for OM theory,''
Nucl. Phys. B \textbf{630}, 178-202 (2002)
doi:10.1016/S0550-3213(02)00175-X
[arXiv:hep-th/0201140 [hep-th]].

\bibitem{GarciaDelMoral:2018jye}
M.~P.~Garcia Del Moral, C.~Las Heras, P.~Leon, J.~M.~Pena and A.~Restuccia,
``M2-branes on a constant flux background,''
Phys. Lett. B \textbf{797}, 134924 (2019)
doi:10.1016/j.physletb.2019.134924
[arXiv:1811.11231 [hep-th]].
\end{thebibliography}
\end{document}